\documentstyle [12pt]{article}
\def\int {\intop \limits}

\def\fnote#1{\footnote}
\begin{document}
\newcommand{\dst}[1]{\displaystyle{#1}}
\newcommand{\barl}{\begin{array}{rl}}
\newcommand{\ball}{\begin{array}{ll}}
\newcommand{\ear}{\end{array}}
\newcommand{\barc}{\begin{array}{c}}
\newcommand{\sne}[1]{\displaystyle{\sum _{#1} }}
\newcommand{\sn}[1]{\displaystyle{\sum ^{\infty }_{#1} }}
\newcommand{\ini}[1]{\displaystyle{\int ^{\infty }_{#1}}}
\newcommand{\myi}[2]{\displaystyle{\int ^{#1}_{#2}}}
\newcommand{\inn}{\displaystyle{\int }}
\newcommand{\be}{\begin{equation}}
\newcommand{\ee}{\end{equation}}
\newcommand{\aq}[1]{\label{#1}}
\renewcommand \theequation{\thesection.\arabic{equation}}

\vspace*{4.0cm}
\centerline{\Large {\bf The Landau-Pomeranchuk-Migdal effect }}
\vskip .25cm
\centerline{\Large {\bf and transition radiation in structured targets}}
\vskip .5cm
\centerline{\large{\bf V. N. Baier and V. M. Katkov}}
\centerline{Budker Institute of Nuclear Physics,
 630090 Novosibirsk, Russia}
\vskip 2.0cm
\begin{abstract}
The radiation from high-energy electrons is investigated
for the case when a target consists of several separated plates.
The spectrum of radiation is considered in the region
in which the bremsstrahlung is under influence of the multiple scattering
of a projectile (the LPM effect), the polarization of a medium
and the hard part of the boundary radiation contribute.
In this region the general expression for the radiation spectrum is
obtained for the $N$-plate target. A qualitative description of the arising
interference pattern is given.

\end{abstract}

\newpage
\section{Introduction}

The process of bremsstrahlung from high-energy electron
occurs over a rather long distance,
known as the formation length.
If the formation length of
the bremsstrahlung becomes comparable to the distance over which
a mean angle of multiple scattering becomes comparable with a
characteristic angle of radiation, the bremsstrahlung will be
suppressed (the Landau-Pomeranchuk-Migdal (LPM) effect \cite{1}, \cite{2}).
An influence of polarization of a medium on radiation process
leads also to suppression of the soft photon emission (Ter-Mikaelian
effect, see in \cite{3}).

A very successful series of experiments \cite{4} - \cite{6} was performed
at SLAC during recent years. In these experiments the cross section
of the bremsstrahlung of soft photons with energy from 200~keV to
500~MeV from electrons with energies 8~GeV and 25~GeV is measured
with an accuracy of the order of a few percent. Both the LPM
effect, and dielectric suppression
(the effects of the polarization of a medium)
were observed and investigated. These experiments were a
challenge for a theory since in all the previous papers calculations
(cited in \cite{7}) were performed to
logarithmic accuracy which is not enough for a description
of the new experiment.

Very recently authors developed the new approach to the theory of the LPM
effect \cite{7}
in which the cross section of the bremsstrahlung process
in the photon energies region where the influence of the LPM is very strong
was calculated with a term $\propto 1/L$ , where $L$
is characteristic logarithm of the problem,
and with the Coulomb corrections
taken into account. In the photon energy region, where the LPM effect
is "turned off", the obtained cross section
gives the exact Bethe-Heitler cross section (within power accuracy) with
the Coulomb corrections. This important feature was absent in
the previous calculations.
The polarization of a medium is incorporated into
this approach. The considerable contribution into the soft part of
the measured spectrum of radiation gives a photon
emission on the boundaries of a target.
In \cite{7} we investigated the case when a target is
much thicker or much thinner than the
formation length of the radiation.
A target of an intermediate thickness was studied in paper \cite{8} .
In the last paper we derived general expression for the spectral
probability of radiation in a thin target and in a target of
intermediate thickness in which the multiple scattering, the
polarization of a medium and radiation on the boundaries of a target
are taken into account. In \cite{8a} the effect of multiphoton emission
from a single electron was studied, this effect is very essential
for understanding data \cite{4}-\cite{6}.
The LPM effect was recently under active
investigation, see e.g. \cite{9}, \cite{10} and review \cite{11}.

In papers \cite{7}, \cite{8} the target is considered as a homogeneous plate.
A radiator which consists of a set of thin plates is of great
interest.

The radiation from several plates for the relatively hard part of the spectrum
in which the bremsstrahlung in the condition of the strong LPM effect
dominates was investigated recently in \cite{12}.A rather
curious interference pattern in the spectrum of the radiation was found
which depends on a number (and a thickness) of plates
and the distance between plates. In this part
of the spectrum one can neglect the effects of the polarization
of a medium.

In the present paper the probability of radiation in a radiator consisting
of $N$ plates is calculated. The transition
radiation dominates in the soft part of the considered spectrum, while
the bremsstrahlung under influence of the strong LPM effect
dominates in the hard part. The intermediate region of the photon energies
where contributions of the both mentioned mechanisms are of the same
order is of evident interest. We consider this region in detail.
In this region effects of the polarization of the medium are essential.
The numerical calculation was performed for the radiator of two gold plates
with thickness $l_1=0.35\%~L_{rad}, L_{rad}$ is the radiation length,
(the same object was considered in \cite{12}).
The interference pattern depending on the distance between plates
was analyzed in the intermediate region.
An another interference pattern was found in the soft part of the spectrum
where the transition radiation contributes only.

\section{Radiation from structured target}

With allowance for the multiple
scattering and the polarization of a medium we
have for the spectral distribution of the probability of radiation
(see Eq.(4.4) of \cite{7},and Eq.(2.1) of \cite{8})
\begin{eqnarray}
&& \frac{dw}{d\omega}= \frac{4\alpha}{\omega}{\rm Re}
\int_{-\infty}^{\infty}dt_2
\int_{-\infty}^{t_2}dt_1 \exp \left(-i\int_{t_1}^{t_2}\mu(t)dt \right)
\nonumber \\
&& \times \left<0|r_1 S(t_2,t_1)+r_2 {\bf p}
S(t_2, t_1) {\bf p}|0 \right>,
\label{1}
\end{eqnarray}
where
\begin{eqnarray}
&& \mu(t)=1+\kappa_0^2 g(t),\quad t=\frac{l}{l_0},
\quad l_0=\frac{2\varepsilon \varepsilon'}{\omega m^2},
\nonumber \\
&& r_1 = \frac{\omega^2}{\varepsilon^2},\quad
r_2=1+\frac{\varepsilon'^2}{\varepsilon^2},\quad
\kappa_0=\frac{\omega_p}{\omega},
\nonumber \\
&& \omega_p=\gamma \omega_0,\quad
\gamma=\frac{\varepsilon}{m}, \quad \omega_0^2=\frac{4\pi \alpha n}{m},
\label{2}
\end{eqnarray}
here $\varepsilon$ is the energy of the initial electron, $\omega$
is the energy of radiated photon, $\varepsilon'=\varepsilon-\omega$,
$n$ is the density of electrons in a medium,
$l$ is the length of the trajectory of a particle,
the function $g(t)$ describes change of the density of a medium on the
trajectory. The mean value in Eq.(\ref{1}) is taken
over states with definite value of the two-dimensional operator
$\mbox{\boldmath$\varrho$}$ (see \cite{7}, Section 2). The propagator
of electron has a form
\begin{equation}
S(t_2, t_1)={\rm T} \exp
\left[-i\int_{t_1}^{t_2} {\cal H}(t) dt \right],
\label{3}\end{equation}
where the Hamiltonian ${\cal H}(t)$ is
\begin{eqnarray}
&& {\cal H} (t)={\bf p}^2-iV(\mbox{\boldmath$\varrho$}) g(t),
\quad {\bf p}=-i\mbox{\boldmath$\nabla$}_{\mbox{\boldmath$\varrho$}},
\nonumber \\
&& V(\mbox{\boldmath$\varrho$})=Q\mbox{\boldmath$\varrho$}^2
\left(L_1+\ln \frac{4}{\mbox{\boldmath$\varrho$}^2}-2C \right), \quad
Q=\frac{2\pi Z^2\alpha^2\varepsilon \varepsilon' n}{m^4\omega},\quad
L_1=\ln \frac{a_{s2}^2}{\lambda_c^2},
\nonumber \\
&& \frac{a_{s2}}{\lambda_c}=183Z^{-1/3}{\rm e}^{-f},\quad
f=f(Z\alpha)=(Z\alpha)^2\sum_{k=1}^{\infty}\frac{1}{k(k^2+(Z\alpha)^2)},
\label{4}
\end{eqnarray}
where $C=0.577216 \ldots$ is Euler's constant.
The contribution of scattering of a projectile on the atomic electrons may
be incorporated into the effective potential $V(\mbox{\boldmath$\varrho$})$.
The summary potential including both an elastic and an inelastic scattering
is
\begin{equation}
V(\mbox{\boldmath$\varrho$})+V_e(\mbox{\boldmath$\varrho$})
=-Q_{ef}\mbox{\boldmath$\varrho$}^2
\Big(\ln \frac{\lambda_c^2}{a_{ef}^2}
+\ln \frac{\mbox{\boldmath$\varrho$}^2}{4}+2C \Big),
\label{4a}\end{equation}
where
\[
\displaystyle{Q_{ef}=Q(1+\frac{1}{Z}),\quad a_{ef}
= a_{s2} \exp \left[\frac{1.88+f(Z \alpha)}{1+Z}\right]}.
\]

In Eq.(\ref{1}) it is implied that the subtraction is made at
$V=0$,~$\kappa=1$.

In \cite{8} the target was one plate of an arbitrary thickness $l_1$
\[
g(t)=\vartheta(t)\vartheta(T_1-t),\quad T_1=l_1/l_0
\]

Here we consider the case when the target consists of N identical plates
of thickness $l_1$ with the equal gaps $l_2$ between them. The case
$l_1 \ll l_c$ will be analyzed where $l_c$ is the characteristic
formation length of radiation in absence of a matter ($\kappa_0 = 0$)
\begin{equation}
l_c=\frac{l_0}{1+\gamma^2\vartheta_c^2}=\frac{\l_0}{1+p_c^2},\quad
\frac{l_1}{l_c}=(1+p_c^2)T_1 \ll 1,
\label{5}
\end{equation}
where $\vartheta_c$ is the characteristic angle of radiation.

In \cite{7} (Sect.5) we obtained the following expression for
the operator $S(t_2, t_1 )$
\begin{equation}
S(t_2, t_1) = \exp(-iH_0t_2)T\exp\left[-\int_{t_1}^{t_2}
V(\mbox{\boldmath$\varrho$}+2{\bf p}t, t) dt\right] \exp(iH_0t_1),
\label{6}\end{equation}
where $H_0={\bf p}^2, V(\mbox{\boldmath$\varrho$}, t) =
V(\mbox{\boldmath$\varrho$}) g(t)$, T means the chronological product.
Let us introduce new variables
\begin{eqnarray}
&& t_1=n_1 T-\tau_1,\quad t_2=n_2T+\tau_2+T_1,\quad T=T_1+T_2, \nonumber \\
&& 0 < \tau_1 < T~(n_1 \geq 1),\quad 0 < \tau_2 < T~(n_2 \leq (N-1)).
\label{7}
\end{eqnarray}
Substituting these variables into Eq.(\ref{6}) we have
\begin{eqnarray}
&& S(t_2, t_1) =\exp\left[-i H_0\left(n_2 T + \tau_2+T_1 \right) \right]
\exp \left[-\int_{n_2 T}^{n_2 T+T_1} V(\mbox{\boldmath$\varrho$}
+2{\bf p}(n_2 T+\tau)) d\tau \right] \nonumber \\
&& \times \exp \left[-\int_{(n_2-1) T}^{(n_2-1) T+T_1}
V(\mbox{\boldmath$\varrho$}+2{\bf p}((n_2-1) T+\tau)) d\tau \right]
\ldots \nonumber \\
&& \times \exp \left[-\int_{n_1 T}^{n_1 T+T_1}
V(\mbox{\boldmath$\varrho$}+2{\bf p}(n_1 T+\tau)) d\tau \right]
\exp \left[iH_0\left(n_1 T + \tau_1 \right) \right]
\label{8}
\end{eqnarray}
Using the equality
\begin{equation}
V(\mbox{\boldmath$\varrho$}+2{\bf p}nT)=\exp(iH_0nT)
V(\mbox{\boldmath$\varrho$})\exp(-iH_0nT)
\label{9}\end{equation}
and the condition (\ref{5}) we obtain
\begin{eqnarray}
&& S(t_2, t_1) \simeq \exp(-iH_0\tau_2)
\exp(-V(\mbox{\boldmath$\varrho$})T_1)\exp(-iH_0T)
\exp(-V(\mbox{\boldmath$\varrho$})T_1)
\nonumber \\
&& \ldots   \exp(-iH_0T)
\exp(-V(\mbox{\boldmath$\varrho$})T_1)\exp(iH_0\tau_1).
\label{10}\end{eqnarray}
Here we neglected the term $-iH_0T_1$ in the exponent
($\exp(-iH_0(\tau_2+T_1))
\newline \rightarrow \exp(-iH_0\tau_2$)) since
$H_0T_1 \sim p_c^2T_1 \ll 1$, we neglected also the term $2{\bf p}\tau$
in the argument of the function $V(\mbox{\boldmath$\varrho$}+2{\bf p}\tau)$
(the term of the order $1/p_c$ is conserved in comparison with the term of
the order $p_cT_1$, see \cite{7}, Sect.5).

We will use below the matrix element of the form
\begin{eqnarray}
&& g_n({\bf p}',{\bf p}) =\left<{\bf p}' \right|
\exp(-a(\mbox{\boldmath$\varrho$}))\exp(-iH_0T)
\exp(-a(\mbox{\boldmath$\varrho$})) \nonumber \\
&& \ldots   \exp(-iH_0T)
\exp(-a(\mbox{\boldmath$\varrho$}))\left|{\bf p} \right>,
\label{11}\end{eqnarray}
where $a(\mbox{\boldmath$\varrho$}) \equiv V(\mbox{\boldmath$\varrho$})T_1$,
the matrix element is calculated between states with definite momentum.
The potential $V(\mbox{\boldmath$\varrho$})$ in Eq.(\ref{4})
we write in the form
\begin{eqnarray}
&& V(\mbox{\boldmath$\varrho$})=V_b(\mbox{\boldmath$\varrho$})
+v(\mbox{\boldmath$\varrho$}),\quad V_b(\mbox{\boldmath$\varrho$})=
q\mbox{\boldmath$\varrho$}^2, \quad q=QL_b,
\nonumber \\
&&L_b \equiv L(\varrho_b)
=\ln \frac{a_{s2}^2}{\lambda_c^2 \varrho_b^2},\quad
v(\mbox{\boldmath$\varrho$})=-\frac{q\mbox{\boldmath$\varrho$}^2}{L}
\left(\ln \frac{\mbox{\boldmath$\varrho$}^2}{4\varrho_b^2}+2C \right),
\label{12}
\end{eqnarray}
where the parameter $\varrho_b$ is defined by a set of equations (we
rearranged terms in
Eq.(5.9) of \cite{7}):
\begin{eqnarray}
&& \varrho_b =1,\quad L_b=L_1\quad {\rm for}\quad 4QL_1T_1=\frac{2\pi}{\alpha}
\frac{l_1}{L_{rad}} \leq 1,\quad q=QL_1;
\nonumber \\
&& 4QL(\varrho_b)T_1\varrho_b^2=\frac{2\pi}{\alpha}
\frac{l_1}{L_{rad}} \varrho_b^2\left(1-\frac{\ln \varrho_b^2}{L_1} \right)
=1,\quad {\rm for}\quad 4QL_1T_1 > 1,
\label{13}
\end{eqnarray}
where $L_1$ is defined in Eq.(\ref{4}).
The parameter $\varrho_b \simeq 1/p_c$ is determined by a characteristic
angle of radiation (momentum transfer).
We will calculate the matrix element $g_n({\bf p}',{\bf p})$
in the first approximation neglecting correction terms containing
$v(\mbox{\boldmath$\varrho$})$. Then
\begin{eqnarray}
&& V(\mbox{\boldmath$\varrho$}) \simeq V_b(\mbox{\boldmath$\varrho$})
=q\mbox{\boldmath$\varrho$}^2, \quad
g_1({\bf p}',{\bf p}) =\left<{\bf p}' \right|
\exp(-qT_1\mbox{\boldmath$\varrho$}^2)
\left|{\bf p} \right>, \nonumber \\
&&=\int_{}^{}d^2\varrho \left<{\bf p}' \right|\mbox{\boldmath$\varrho$}
\left> \right< \mbox{\boldmath$\varrho$} \left|
\exp(-qT_1\mbox{\boldmath$\varrho$}^2) \right| \mbox{\boldmath$\varrho$}
\left> \right<\mbox{\boldmath$\varrho$}\left|{\bf p} \right> \nonumber \\
&&=\frac{1}{(2\pi)^2}\int_{}^{}d^2\varrho \exp\left[i({\bf p}
-{\bf p}')\mbox{\boldmath$\varrho$} -qT_1\mbox{\boldmath$\varrho$}^2\right]
=\frac{b}{\pi}\exp \left[-b\left({\bf p}
-{\bf p}' \right)^2 \right],
\label{14}\end{eqnarray}
where $b=1/(4qT_1)$.
In the calculation of the expression (\ref{11}) we insert the combinations of
the state vectors $\left|{\bf p}_1 \right> \left<{\bf p}_1 \right| \ldots
\left|{\bf p}_{n-1} \right> \left<{\bf p}_{n-1} \right|$ between the operators
$\exp (-a(\mbox{\boldmath$\varrho$}))$ and use the matrix element
calculated in (\ref{14}), we have
\begin{eqnarray}
&& g_n({\bf p}',{\bf p}) = \left(\frac{b}{\pi}\right)^n
\int_{}^{} d^2p_1 \int_{}^{} d^2p_2 \ldots \int_{}^{} d^2p_{n-1} \nonumber \\
&& \exp \left[-b\left({\bf p}'-{\bf p}_1 \right)^2 -i{\bf p}_1^2T
-b\left({\bf p}_1-{\bf p}_2 \right)^2 -i{\bf p}_2^2T \ldots
-b\left({\bf p}_{n-1}-{\bf p} \right)^2 -i{\bf p}_{n-1}^2T\right] \nonumber \\
&& =\frac{b}{\pi d_n} \exp \left\{-\frac{b}{d_n} \left[
\left({\bf p}'^2+{\bf p}^2 \right)\left(d_n-d_{n-1} \right)-
2{\bf p}{\bf p}'\right]\right\},
\label{15}\end{eqnarray}
where
\begin{equation}
d_n \equiv d_n(\alpha) = \frac{1}{2\sqrt{\alpha^2-1}}
\left[\left(\alpha+\sqrt{\alpha^2-1} \right)^n-
\left(\alpha-\sqrt{\alpha^2-1} \right)^n \right],\quad \alpha=1+\frac{iT}{2b}
\label{16}\end{equation}
The obtained function $d_n(\alpha)$ is the polynomial with respect to
$\alpha$ of degree $(n-1)$
\begin{equation}
d_0=0,\quad d_1=1, \quad d_2=2\alpha, \quad d_3=4\alpha^2-1,
\quad d_4=4\alpha \left(2\alpha^2-1 \right) \ldots
\label{17}\end{equation}
After substitution $\alpha=\cosh \eta$ we find from Eqs.(\ref{16})
and (\ref{15})
\begin{eqnarray}
&& d_n=\frac{\sinh n\eta}{\sinh \eta},\quad
g_n({\bf p}',{\bf p}) = \frac{b \sinh \eta}{\pi \sinh n\eta}
\exp \Bigg\{-\frac{b}{\sinh n\eta}  \nonumber \\
&& \times \left[
\left({\bf p}'^2+{\bf p}^2 \right)\left(\sinh n\eta -\sinh (n-1)\eta \right)-
2 \sinh \eta{\bf p}{\bf p}'\right]\Bigg\},
\label{18}\end{eqnarray}
where
\[
\sinh \eta = 2\sqrt{iqT_1T\left(1+iqT_1T \right)}.
\]
For the case $\eta \ll 1$ one has
\begin{eqnarray}
&& \eta \simeq 2\sqrt{iqT_1T} \equiv 2T\sqrt{i\overline{q}},\quad
\overline{q}=q\frac{T_1}{T}=q\frac{l_1}{l_1+l_2}, \nonumber \\
&& g_n({\bf p}',{\bf p}) \simeq \frac{b \eta}{\pi \sinh n\eta}
\exp \left\{-\frac{b \eta}{\sinh n\eta}
\left[\left({\bf p}'^2+{\bf p}^2 \right)\cosh n\eta -
2 {\bf p}{\bf p}'\right] \right\}.
\label{19}\end{eqnarray}
For the limiting case $n\eta \ll 1$ one has
\begin{equation}
g_n({\bf p}',{\bf p}) \simeq \frac{b}{\pi n}
\exp \left\{-\frac{b}{n}
\left({\bf p}'-{\bf p}\right)^2\right\}.
\label{20}\end{equation}

Using the results obtained we can calculate now the mean value
entering Eq.(\ref{1})
\begin{eqnarray}
&& \left<0 \right| S_n(t_2, t_1)\left|0 \right> = \int_{}^{}d^2p'
\int_{}^{}d^2p \left<0 \right|{\bf p}' \left> \right<
{\bf p'} \left| S_n(t_2, t_1)\right|0 \left> \right<{\bf p}\left|0 \right>
\nonumber \\
&&=\frac{1}{(2\pi)^2}\int_{}^{}d^2p' \int_{}^{}d^2p
\exp \left(-i{\bf p}'^2\tau_2-i{\bf p}^2\tau_1 \right)
g_n({\bf p}',{\bf p})=\frac{b}{4\pi d_n \gamma_n}, \nonumber \\
&& \gamma_n=\sigma_n^2-\beta_n^2+i\sigma_n(\tau_1+\tau_2)-\tau_1\tau_2,\quad
\beta_n=\frac{b}{d_n},\quad \sigma_n=\beta_n(d_n-d_{n-1}).
\label{21}\end{eqnarray}
If we use the same procedure for the second mean value in (\ref{1})
we obtain
\begin{equation}
\left<0 \right|{\bf p} S_n(t_2, t_1) {\bf p}\left|0 \right>=
\frac{b^2}{4\pi d_n^2 \gamma_n^2}.
\label{22}\end{equation}

We split now the spectral distribution of the probability of radiation
into two parts
\begin{eqnarray}
&& \frac{dw}{d\omega} = \frac{dw_{br}}{d\omega} + \frac{dw_{tr}}{d\omega};
\nonumber \\
&& \frac{dw_{br}}{d\omega}= \frac{4\alpha}{\omega}{\rm Re}
\int_{-\infty}^{\infty}dt_2
\int_{-\infty}^{t_2}dt_1 \exp \left(-i\int_{t_1}^{t_2}\mu(t)dt \right)
\nonumber \\
&& \times \left<0 \right|r_1 \left( S(t_2,t_1)-S^{(0)}(t_2,t_1) \right)+
r_2 {\bf p}\left( S(t_2,t_1)-S^{(0)}(t_2,t_1) \right) {\bf p}\left|0 \right>.
\label{23}\end{eqnarray}
\begin{eqnarray}
&& \frac{dw_{tr}}{d\omega}= \frac{4\alpha}{\omega}{\rm Re}
\int_{-\infty}^{\infty}dt_2
\int_{-\infty}^{t_2}dt_1 \left[\exp \left(-i\int_{t_1}^{t_2}\mu(t)dt \right)
-\exp \left(-i(t_2-t_1) \right) \right] \nonumber \\
&& \times \left<0 \right|r_1 S^{(0)}(t_2,t_1) +
r_2 {\bf p}S^{(0)}(t_2,t_1) {\bf p}\left|0 \right>, \nonumber \\
&& S^{(0)}(t_2,t_1) = \exp \left[-iH_0\left(t_2-t_1 \right) \right],
\label{24}\end{eqnarray}
where $dw_{br}/d\omega$ is the spectral distribution of the
probability of bremsstrahlung with allowance for the multiple
scattering and polarization of a medium, $dw_{tr}/d\omega$ is
the probability of the transition radiation obtained in the frame of
quantum electrodynamics.

Note that the subtraction in Eq.(\ref{23}) has to follow the
procedure
\begin{equation}
\left<0 \right| S_n\left|0 \right>-\left<0 \right| S_n(q=0)\left|0 \right>.
\label{25}\end{equation}

The integral in the exponential in (\ref{23}) is
\begin{equation}
\int_{t_1}^{t_2} \mu(t) dt =\tau_1+\tau_2+\kappa T_1+(n-1)\overline{\kappa}T,
~\kappa=1+\kappa_0^2,~\overline{\kappa}=1+\overline{\kappa_0^2},
~\overline{\kappa_0^2}=\frac{\kappa_0^2 l_1}{l_1+l_2},
\label{26}\end{equation}
where we used notations introduced in (\ref{2}) and (\ref{7}).

Substituting (\ref{21}), (\ref{22}), (\ref{25}) and (\ref{26})
into Eq.(\ref{23}) we obtain the following expression for
the spectral distribution of the
probability of the bremsstrahlung with allowance for the multiple
scattering and the polarization of a medium for the $N$-plate target
\begin{eqnarray}
&& \frac{dw_{br}^{(N)}}{d\omega}= \frac{\alpha}{\pi \omega}{\rm Re}
\sum_{n_1,n_2=0}^{N-1}\int_{}^{}d\tau_2
\int_{}^{}d\tau_1 \exp \left[-i\left(\tau_1+\tau_2+\kappa T_1+
(n-1)\overline{\kappa}T \right) \right]
\nonumber \\
&& \times \left[r_1 \left(G_n-G_n(0) \right)+
r_2 \left(G_n^2-G_n^2(0)  \right)  \right],
\label{27}\end{eqnarray}
where
\begin{eqnarray}
&& G_n^{-1}= \beta_n\left[\left(d_n-d_{n-1} \right)^2-1 \right]+
i\left(d_n-d_{n-1} \right)\left(\tau_1+\tau_2\right)-
\beta_n^{-1}\tau_1\tau_2; \nonumber \\
&& N \geq n=n_2-n_1+1 \geq 1;\quad n_1=0,~0 \leq \tau_1 < \infty; \quad
n_2=N-1,~0 \leq \tau_2 < \infty; \nonumber \\
&&  n_1 \geq 1,~ ~0 \leq \tau_1 \leq T;\quad
n_2 \leq N-2,~0 \leq \tau_2 \leq T,
\label{28}\end{eqnarray}
here we used (\ref{21}), (\ref{22}) and (\ref{7}). Note for the subtraction
procedure one has that when
$q \rightarrow 0$, the function $b,~\beta_n \rightarrow \infty$.

The formula (\ref{27}) can be rewritten in the form which is more
convenient for application
\begin{eqnarray}
&& \frac{dw_{br}^{(N)}}{d\omega}= \frac{\alpha}{\pi \omega}{\rm Re}
\int_{0}^{\infty}d\tau_2
\int_{0}^{\infty}d\tau_1 \exp \left[-i\left(\tau_1+\tau_2+\kappa T_1
 \right) \right] \Bigg\{\sum_{n=1}^{N-1}
\exp \left[-i(n-1)\overline{\kappa}T \right]
\nonumber \\
&& \times R_n(\tau_1, \tau_2) \left[\vartheta(T-\tau_1) +
\vartheta(T-\tau_2) +(N-n-1)\vartheta(T-\tau_1)\vartheta(T-\tau_2) \right]+
\nonumber \\
&& \exp \left[-i(N-1)\overline{\kappa}T \right]
R_N(\tau_1, \tau_2) \Bigg\},
\label{27a}
\end{eqnarray}
where
\begin{equation}
R_n(\tau_1, \tau_2) = r_1 \left(G_n-G_n(0) \right)+
r_2 \left(G_n^2-G_n^2(0)  \right).
\label{28a}
\end{equation}

For one plate ($N=1$) we have $n_1=n_2=0,~n=1$ and
\begin{equation}
G_1^{-1}=i(\tau_1+\tau_2)-\frac{1}{b}\tau_1\tau_2,\quad
G_1^{-1}(0)=i(\tau_1+\tau_2).
\label{29}\end{equation}
In the integral (\ref{27}) we rotate the integration contours over
$\tau_1,~\tau_2$ on the angle $-\pi/2$ and substitute variables
$\tau_{1,2} \rightarrow -ix_{1,2}$. Then we carry out change of
variables $x=x_1+x_2,~x_2=zx$. We have
\begin{eqnarray}
\hspace{-8mm}&& \frac{dw_{br}^{(1)}}{d\omega}=
\frac{\alpha}{\pi \omega} \cos \kappa T_1
\int_{0}^{\infty} \exp (-x)dx\int_{0}^{1}\left[r_1\left(1-
\frac{1}{g(x,z)} \right)+r_2\left(1-\frac{1}{g^2(x,z)} \right) \right]dz
\nonumber \\
\hspace{-8mm}&& = \frac{\alpha b}{\pi \omega}
\cos (\kappa T_1) \int_{0}^{\infty}
\exp (-bx) \left[r_1F_1\left(\sqrt{\frac{x}{4}} \right) +
r_2F_2\left(\sqrt{\frac{x}{4}} \right)\right],\quad
g(x,z)=1+\frac{x}{b}z(1-z), \nonumber \\
\hspace{-8mm}&& F_1(u)=1-
\frac{\ln \left(u+\sqrt{1+u^2} \right)}{u\sqrt{1+u^2}},\quad
F_2(u)=\frac{1+2u^2}{u\sqrt{1+u^2}}\ln \left(u+\sqrt{1+u^2} \right)-1.
\label{30}\end{eqnarray}
Here the integration by parts is carried out in the term $\propto r_2$
first over $x$ and then over $z$ in the term containing
$\ln (1+xz(1-z))$.
The expression (\ref{30}) was derived in \cite{7}, Sect.5 (see also
references therein) with allowance for the correction term
$v(\mbox{\boldmath$\varrho$})$ (see Eq.(\ref{12})).

We handle now to the case when a target consists of two plates ($N=2$).
Since the formation length is enough long for the soft photons
($\omega \ll \varepsilon$) only, we consider the term with $r_2$ in
(\ref{27}) as far as $r_1=\omega^2/\varepsilon^2 \ll 1$.
For the case ($N=2$) the sum in (\ref{27}) consists of three terms:
$1) n_1=n_2=0;~2) n_1=n_2=1;~3) n_1=0, n_2=1$. For the two first $n=1$ and
we have from (\ref{29})
\begin{eqnarray}
&& \frac{dw_{br1}^{(2)}}{d\omega}= \frac{dw_{br2}^{(2)}}{d\omega}=
\frac{\alpha r_2}{\pi \omega}~ {\rm Re}~\Bigg[ \exp (-i\kappa T_1)
\int_{0}^{\infty} d\tau_2\int_{0}^{T} \exp (-i(\tau_1+\tau_2))
\nonumber \\
&& \times \left[\frac{1}{(\tau_1+\tau_2)^2}-
\frac{1}{(\tau_1+\tau_2+i\tau_1\tau_2/b)^2} \right]d\tau_1 \Bigg].
\label{31}\end{eqnarray}
For the third term $n=2,~d_2=2\alpha$ (see (\ref{17})) and we have
\begin{eqnarray}
&& G_2^{-1}=iT+i\left(1+\frac{iT}{b} \right)(\tau_1+\tau_2)
-\frac{2}{b}\left(1+\frac{iT}{2b} \right)\tau_1\tau_2
\nonumber \\
&& = i(T+\tau_1+\tau_2)-\frac{2}{b}\tau_1\tau_2 +\frac{iT}{b}
\left[i(\tau_1+\tau_2)-\frac{\tau_1\tau_2}{b} \right],
\label{32}\end{eqnarray}
so that
\begin{eqnarray}
\hspace{-14mm}&& \frac{dw_{br3}^{(2)}}{d\omega}=
\frac{\alpha r_2}{\pi \omega}~ {\rm Re}
~\Bigg[ \exp (-i(\overline{\kappa} T+\kappa T_1))
\int_{0}^{\infty} dx_1\int_{0}^{\infty} \exp (-(x_1+x_2))
\nonumber \\
\hspace{-14mm}&& \times \left[\frac{1}{(iT+x_1+x_2)^2}-
\frac{1}{\left[iT+x_1+x_2+\frac{2}{b}x_1x_2 +\frac{iT}{b}
\left(x_1+x_2+\frac{x_1x_2}{b}  \right)\right]^2} \right]dx_2 \Bigg].
\label{33}\end{eqnarray}
Here we rotate the integration contours over
$\tau_1,~\tau_2$ on the angle $-\pi/2$ and substitute variables
$\tau_{1,2} \rightarrow -ix_{1,2}$.

In the case of the weak multiple scattering ($b \gg 1$)  neglecting
the effect of the polarization of a medium ($\kappa = 1$) and expanding
the integrand in (\ref{31}) and (\ref{33}) over $1/b$ we
have for the probability of radiation
\begin{eqnarray}
&& \frac{dw_{br}^{(2)}}{d\omega}=2 \frac{dw_{br1}^{(2)}}{d\omega}+
\frac{dw_{br3}^{(2)}}{d\omega} \simeq \frac{2\alpha r_2}{3\pi \omega b}~
\left[1-\frac{3}{10b}+\frac{1}{b}G(T) \right],
\nonumber \\
&& G(T)=T^2 \int_{0}^{\infty} \frac{x^3 \exp (-x)}{(x^2+T^2)^2}
\Bigg[\left(1-\frac{3x^2}{10T^2} \right)
\cos\left(T-4\arctan\frac{x}{T} \right)
\nonumber \\
&& +\frac{2x}{T}
\sin\left(T-4\arctan\frac{x}{T} \right) \Bigg].
\label{34}
\end{eqnarray}
If the distance between plates is small ($T \ll 1$) we have
\begin{equation}
G(T) \simeq -\frac{3}{10}+12T^2\left(\ln \frac{1}{T}-C-\frac{19}{24} \right).
\label{35}
\end{equation}
In the opposite case ($T \gg 1$) one can
obtain asymptotic expansion of $G(T)$ using the method of stationary phase
(similar method wasw used in derivation of the Stirling formula)
\begin{equation}
G(T) \simeq \frac{6T^2}{(9+T^2)^2}\cos\left(T-4\arctan\frac{3}{T} \right)
+\frac{48T}{(16+T^2)^2}\sin\left(T-4\arctan\frac{4}{T} \right).
\label{36}
\end{equation}
Note that the main term of the decomposition in Eq.(\ref{34}) is
the Bethe-Heitler probability of radiation from two plates which is
independent of the distance between plates. This means that in the case
considered  we have independent radiation from each plate
without interference in the main order over $1/b$. The interference effects
appear only in the next orders over $1/b$.

In the case of the strong multiple scattering ($b \ll 1,~p_c^2 \gg 1$)
we consider first
the transition region from $T \ll b$ (the formation length is much longer
than the distance between plates) to $T \gg b$ (at small $T, T \ll 1$).
We introduce parameter $\delta = iT/b$, then
assuming $b \ll 1, T \ll 1$ we obtain
\begin{eqnarray}
&& \frac{dw_{br1}^{(2)}}{d\omega}= \frac{dw_{br2}^{(2)}}{d\omega}=
\frac{\alpha r_2}{\pi \omega}~ {\rm Re}~\Big[ \exp (-i\kappa T_1)
f_1(\delta) \Big],
\nonumber \\
&& \frac{dw_{br3}^{(2)}}{d\omega}=
\frac{\alpha r_2}{\pi \omega}~ {\rm Re}~\Big[ \exp (-2i\kappa T_1)
f_2(\delta) \Big],
\label{38}\end{eqnarray}
where
\begin{eqnarray}
&& f_1(\delta)=(1+b)\ln (1+\delta)+\frac{b\delta^2}{1+\delta}
\left(\ln (b\delta)+C-1 \right)+b\delta\left(\frac{\ln (1+\delta)}{1+\delta}
-1\right),
\nonumber \\
&& f_2(\delta)=\ln \left(\frac{2+\delta}{b(1+\delta)^2} \right)-1-C +
b\left[\ln \left(\frac{2+\delta}{b(1+\delta)}\right)+1-C \right]
-\frac{b\delta}{1+\delta} \Big[ (1+2\delta) \nonumber \\
&& \times (\ln (b\delta)+C)+\ln (1+\delta) -\delta \Big] +
\frac{b\delta}{(1+\delta)(2+\delta)}
\ln \left(\frac{b\delta(1+\delta)}{2+\delta} \right).
\label{39}
\end{eqnarray}
When the formation length is much larger than the target thickness
as a whole ($T \ll b,~|\delta| \ll 1$), one can decompose
the functions $f_1(\delta)$ and  $f_2(\delta)$ into the Taylor series over
$\delta$. Retaining the main terms of the expansion we find
for the probability of radiation
\begin{equation}
\frac{dw_{br}^{(2)}}{d\omega} \simeq \frac{\alpha r_2}{\pi \omega}
\left[(1+b)\left(\ln \frac{2}{b}+1- C\right)-2 \right] \cos (2\kappa T_1)
\label{40}
\end{equation}
In the opposite case $b \ll T \ll 1~(|\delta| \gg 1)$
neglecting the effect of the polarization of a medium
($\kappa T_1 \ll 1$) we have
\begin{equation}
\frac{dw_{br}^{(2)}}{d\omega} \simeq \frac{\alpha r_2}{\pi \omega}
\left[(1+2b)\left(\ln \frac{T}{b^2}+1- C\right)-2 \right]
\label{41}
\end{equation}
Note that when $T=1$ the probability (\ref{41}) is within logarithmic accuracy
doubled probability of radiation from one plate with the thickness $l_1$:
\begin{equation}
\frac{dw_{br}^{(1)}}{d\omega} \simeq \frac{\alpha r_2}{\pi \omega}
\left[(1+2b)\left(\ln \frac{1}{b}+1- C\right)-2 \right]
\label{42}
\end{equation}

The case of the strong multiple scattering ($b \ll 1$) for the
photon energies where the value $T \geq 1$ is of the special interest.
In this case we can neglect the polarization of a medium
$\kappa=\overline{\kappa}=1$, and disregard the terms $\propto \kappa T_1$
in the exponent of the expressions (\ref{31}) and (\ref{33}) since
$T_1 \ll 1$. In the integral over $\tau_1$ in (\ref{31}) we add and subtract
the contribution of the interval $T \leq \tau_1 < \infty$. The sum gives
Eq.(\ref{30}), i.e. the radiation from one plate, and in the difference
the main contribution gives region $\tau_2 \sim 1$ so that one can
disregard the second terms in the square brackets in Eqs.(\ref{31}) and
(\ref{33}). We obtain as a result
\begin{eqnarray}
&& \frac{dw_{br}^{(2)}}{d\omega}=2\frac{dw_{br1}^{(2)}}{d\omega}+
\frac{dw_{br3}^{(2)}}{d\omega} \simeq 2\frac{dw_{br}^{(2)}}{d\omega}-
\frac{dw_{br3}^{(2)}}{d\omega},\quad \frac{dw_{br3}^{(2)}}{d\omega}=
\frac{\alpha r_2}{\pi \omega}~F(T),
\nonumber \\
&&{\rm Re}~F(T)=\int_{0}^{\infty}d\tau_2 \int_{T}^{\infty} d\tau_1
\frac{1}{(\tau_1+\tau_2)^2} \exp(-i(\tau_1+\tau_2))=
\int_{T}^{\infty} \frac{d\tau}{\tau^2}\exp(-i\tau) (\tau-T)
\nonumber \\
&&=-({\rm ci}(T) + T{\rm si}(T) + \cos T),
\label{42a}
\end{eqnarray}
where ${\rm si}(z)$ is the integral sine and ${\rm ci}(z)$
is the integral cosine.
At $T \gg 1$ we have
\begin{equation}
F(T) = -\frac{\cos T}{T^2}.
\label{42b}
\end{equation}

We carried out the analysis of cases $N=1,2$ using Eq.(\ref{27}).
We illustrate an application of Eq.(\ref{27a}) for the case $N=4$:
\begin{eqnarray}
\hspace{-8mm}&& \frac{dw_{br}^{(4)}}{d\omega}=
\frac{\alpha}{\pi \omega}{\rm Re} \Bigg\{\exp (-i\kappa T_1) \Bigg[
2\int_{0}^{\infty}d\tau_2
\int_{0}^{T}d\tau_1 \exp \left[-i\left(\tau_1+\tau_2\right) \right]
\Big(R_1+\exp \left(-i\overline{\kappa}T\right) R_2
\nonumber \\
\hspace{-8mm}&& +\exp \left(-2i\overline{\kappa}T\right) R_3 \Big) +
\int_{0}^{T}d\tau_2
\int_{0}^{T}d\tau_1 \exp \left[-i\left(\tau_1+\tau_2\right) \right]
\left(2R_1+\exp \left(-i\overline{\kappa}T\right) R_2  \right)
\nonumber \\
\hspace{-8mm}&& + \exp \left(-i3\overline{\kappa}T \right)
\int_{0}^{\infty}d\tau_2
\int_{0}^{\infty}d\tau_1 \exp \left[-i\left(\tau_1+\tau_2\right) \right]
R_4  \Bigg] \Bigg\}.
\label{42c}
\end{eqnarray}

We consider now the case of large $N$ ($N \geq 3$).
If the formation length of
the bremsstrahlung is shorter than the distance between plates ($T > 1$)
the interference of the radiation from neighboring plates takes place.
Using the probability of radiation from two plates (\ref{34}) we obtain
in the case of weak multiple scattering ($b \gg 1$)
\begin{equation}
\frac{dw_{br}^{(N)}}{d\omega}
\simeq \frac{N\alpha r_2}{3\pi \omega b}~
\left[1-\frac{3}{10b}+2\frac{N-1}{Nb}G(T) \right],
\label{43}
\end{equation}
where for $T \gg 1$ the function $G(T)$ is defined in (\ref{34}).
In the case of strong multiple scattering ($b \ll 1$) and large $T$
we have (compare with Eqs.(\ref{42a}) and (\ref{42b}))
\begin{equation}
\frac{dw_{br}^{(N)}}{d\omega}
\simeq N\left(\frac{dw_{br}^{(1)}}{d\omega} +
\frac{\alpha r_2}{\pi \omega b}~
\frac{N-1}{Nb}\frac{\cos T}{T^2} \right).
\label{44}
\end{equation}

In the opposite limiting case $|\eta| \ll 1,~(\eta^2 \simeq \delta)
~N|\eta| \ll 1$
(see Eqs.(\ref{19}) and (\ref{20})), i.e. when the formation length
of the bremsstrahlung is longer than the radiator thickness,
the radiation act takes
place on a target as a whole. In this case, as it follows from Eq.
(\ref{20}), the parameter $b$ diminishes $N$ times (the value $p_c^2$
increases $N$ times). The analysis of the case of two plates conducted above
in detail supports this result. In the limiting case of strong multiple
scattering ($b \ll 1$) one can see this from (\ref{40}).

In the case $|\eta| \ll 1$ (the formation length
of the bremsstrahlung is longer than a distance between plates as before)
and large $N$ (including the case when $|N|\eta| \gg 1$)
one can substitute the summation over $n$ by integration
in the expression for the probability of radiation (\ref{27}). Using
Eqs.(\ref{18})-(\ref{21}) we have
\begin{eqnarray}
&& n\eta \simeq nT2\sqrt{i\overline{q}} \rightarrow t\nu,\quad
b\eta \simeq \sqrt{ibT}=\frac{i}{2\sqrt{i\overline{q}}}=\frac{i}{\nu},\quad
d_n \rightarrow \frac{\sinh \nu t}{\eta},
\nonumber \\
&& \beta_n \rightarrow \frac{b\eta}{\sinh \nu t}=\frac{i}{\nu \sinh \nu t},
\quad d_n-d_{n-1} \rightarrow \cosh \nu t;
\nonumber \\
&& G_n^{-1} \rightarrow \frac{i}{\nu}\left[\sinh \nu t
+\nu\left(\tau_1+\tau_2 \right)\cosh \nu t + \nu^2 \tau_1 \tau_2
\sinh \nu t\right],
\label{45}
\end{eqnarray}
where $\nu=2\sqrt{i\overline{q}}$ (see \cite{7}, Sect.2).
The four regions contribute into the sum and integrals
over $\tau_1,\tau_2$ in Eq.(\ref{27}), see also Eq.(\ref{27a}).
\begin{enumerate}
\item The first region $0 \leq \tau_1 < \infty,~ 0 \leq \tau_2 \leq T~
(\tau_1 \rightarrow t_1,~nT \rightarrow t_2)$, so we have in this region
\[
G_n \rightarrow -iN_1,\quad N_1 \simeq \frac{\nu}{\left(\sinh \nu t_2
+\nu t_1\cosh \nu t_2  \right)},
\]
where we take into account that $\nu T=\eta \ll 1$.
\item In the second region $0 \leq \tau_1,\tau_2 \leq T~(nT \rightarrow
t_2-t_1=t)$
\[
G_n \rightarrow -iN_2,\quad N_2 \simeq \frac{\nu}{\sinh \nu t}.
\]
\item The third region gives the same contribution after substitution
$\tau_1 \leftrightarrow \tau_2,~t_1 \leftrightarrow t_2$.
\item  In the fourth region $0 \leq \tau_1,\tau_2 < \infty~(\tau_{1,2}
\rightarrow t_{1,2})$
\[
G_n \rightarrow -iN_4,\quad
N_4 \simeq \frac{\nu}{(1+\nu^2t_1t_2)\sinh \nu NT
+\nu (t_1+t_2)\cosh \nu NT}.
\]
\end{enumerate}
Substituting these expressions into Eq.(\ref{27}) we arrive to the
formula (2.11) of \cite{8} which describes radiation on the plate of
the thickness $NT$ (in units of the formation length). This case was
analyzed in detail in \cite{8}.

\section{A qualitative analysis of the radiation in the structured target}

\setcounter{equation}{0}

We investigate the behavior of the spectral distribution
$\displaystyle{\omega \frac{dw}{d\omega}}$ using as an example the case
of two plates with the thickness $l_1$ and the distance between plates
$l_2 \geq l_1$ which was analyzed in detail in the previous Section.
For plates with the thickness $l_1 \geq 0.2\% L_{rad}$
and in the energy interval $\omega > \omega_p$, in which the
effects of the polarization of a medium can be discarded,
the condition (\ref{6}) is fulfilled only for enough high energy
$\varepsilon$, when the characteristic energy
\begin{equation}
\displaystyle{\omega_c=\frac{16\pi Z^2 \alpha^2}{m^2}
\gamma^2 n_a \ln \frac{a_{s2}}{\lambda_c}},
\label{46}
\end{equation}
where $n_a$ is the number density of atoms in the medium,
is such that $\omega_p \ll \omega_c$.
We study the situation
when the LPM suppression of the intensity of radiation
takes place for relatively soft energies of photons:
$\omega \leq \omega_c \ll \varepsilon$.

We consider first the hard photons
$\omega_c \ll \omega < \varepsilon$. In this interval of $\omega$
the formation length $l_0$ (\ref{2}) is much shorter than the plate
thickness $l_1~(T_1 \gg 1)$, the radiation intensity is the incoherent
sum of radiation from two plates and it is independent of the distance
between plates. In this interval the Bethe-Heitler formula is valid.

For $\omega \leq \omega_c$ the LPM effect turns on, but when
$\omega = \omega_c$ the thickness of plate is still larger
than the formation length $l_0$ (the opposite case will be considered
in the end of the Section)
\begin{equation}
\frac{l_1}{l_0(\omega_c)} = T_1(\omega_c) \equiv T_c
\simeq \frac{2\pi}{\alpha}\frac{l_1}{L_{rad}} > 1,
\label{47}
\end{equation}
so that the formation of radiation takes place mainly inside each of plates.
With $\omega$ decreasing we get over to the region where
the formation length $l_c > l_1$, but effects of the polarization
of a medium are still weak ($\omega > \omega_p$).
Within this interval (for $\omega < \omega_{th}$) the main condition
(\ref{5}) is fulfilled. To estimate the value $\omega_{th}$ we have to take
into account the characteristic radiation angles ($p_c^2$ in Eq.(\ref{5})),
connected with mean square angle of the multiple scattering. Using
Eq.(\ref{13})) and the definition of the parameter $b=1/(4qT_1)$ in
Eq.(\ref{14})) we find
\begin{eqnarray}
&& p_c^2 \leq \frac{1}{b}=\frac{2\pi}{\alpha}\frac{l_1}{L_{rad}}
\left(1-\frac{\ln \varrho_b}{\ln (a_{s2}/\lambda_c)} \right)
\simeq T_c \left(1+\frac{\ln T_c}{2\ln (a_{s2}/\lambda_c)} \right);
\nonumber \\
&& \omega_{th}=\frac{\omega_c}{T_c\left(1+p_c^2 \right)}
\geq \omega_b=\frac{\omega_c}{T_c(1+T_c)}.
\label{48}
\end{eqnarray}
It is shown in \cite{8} (Sec.3, see discussion after Eq.(3.6))
that $\omega_{th} \simeq 4\omega_b$ (in \cite{8} the notation
$\omega_2$ was used instead of $\omega_b$). Naturally,  $\omega_{th}
< \omega_c/T_c$, and when $\omega =\omega_c/T_c$ one has $T_1=1,~
l_1=l_0$. It is seen from Eq.(\ref{48}) that when the value $l_1$
decreases, the region of applicability of results of this paper grows.

So, when $\omega < \omega_{th}$ the formation length is longer
than the thickness of the plate $l_1$ and the coherent effects
depending on the distance between plates $l_2$ turn on.
For the description of these effects for $T=\left(1+l_2/l_1 \right) T_1
\geq 1$ one can use Eq.(\ref{42a}). For $T \geq \pi \gg 1$
one can use the asymptotic expansion (\ref{42b}) and it is seen
that at $T=\pi$ the spectral curve has minimum. Let us note an accuracy
of formulas is better when $\omega$ decreases, and the description is
more accurate for $T \gg T_1~(l_2 \gg l_1)$.

With further decreasing of the photon energy $\omega$ the value $T$
diminishes and the spectral curve grows until $T \sim 1$. When $T < 1$
the spectral curve decreases $\propto \ln T$
according with the second formula
of Eq.(\ref{42}). So, the spectral curve has maximum for $T \sim 1$.
The mentioned decreasing continues until the photon energy $\omega$
for which $(1+2/b)T \sim 1$. For smaller $\omega$ the thickness of the target
is shorter than the formation length. In this case the first
Eq.(\ref{42}) is valid which is independent of the value $T$.

The next characteristic region of the photon energies is $\omega \leq
\omega_p$ where the polarization of a medium is manifest itself.
For $\omega \sim \omega_0^2 l_1 \ll \omega_p$ one has $\kappa T_1 \sim 1$
and for the bremsstrahlung contribution instead of Eq.(\ref{42})
we have to use Eqs.(\ref{40})-(\ref{41}) which include the interference
of the bremsstrahlung on the plate boundaries. However, in this region
the transition radiation gives the main contribution.

The spectral probability of transition radiation in the radiator
consisting of $N$ thin plates of the thickness $l_1$ separated by equal
distances $l_2$ was discussed in many papers,
see e.g. \cite{3}. It has the form
\begin{equation}
\frac{dw_{tr}^{(N)}}{d\omega}=\frac{4 \alpha}{\pi \omega}
\int_{0}^{\infty} \frac{y dy}{(1+y)^2}
\left(\frac{\kappa_0^2}{(1+\kappa_0^2+y)} \right)^2
\Phi_N(y),
\label{49}
\end{equation}
where
\begin{equation}
\Phi_N(y)=
\sin^2\frac{\varphi_1}{2}~\frac{\sin^2 (N\varphi/2)}{\sin^2 (\varphi/2)},
\label{50}
\end{equation}
here
$y=\vartheta^2 \gamma^2$, $\vartheta$ is the angle
of emission with respect velocity of the incident electron (we assume
normal incidence) and
\begin{eqnarray}
&& \varphi_1=\frac{\omega l_1}{2\gamma^2}
\left(1+\kappa_0^2+y \right) \simeq T_1(\kappa+y),
\nonumber \\
&&\varphi=\frac{\omega l}{2\gamma^2}
\left(1+y \right) +\frac{\omega l_1}{2\gamma^2} \kappa_0^2
\simeq T(1+y)+\kappa T_1
\label{51}
\end{eqnarray}
The formula (\ref{49}) can be derived directly from Eq.(\ref{24}) if
one gets over to the $p$-representation (make
\[
\int_{}^{}d^2p|{\bf p}><{\bf p}|, \quad |<{\bf p}|0>|^2=\frac{1}{(2\pi)^2}),
\]
than one substitutes the real part of the double integral over time by
one-half of the modulus squared of the single integral. After substitution
${\bf p}^2=y,~d^2p=\pi dp^2=\pi dy$ we pass to Eq.(\ref{49}).

In the case $N=2$ one has
\begin{equation}
\Phi_2(y)=
4\sin^2\frac{\varphi_1}{2}~\cos^2 \frac{\varphi}{2}.
\label{52}
\end{equation}
In the integral in (\ref{49}) for $y < 2/T \gg 1$ the function
$\Phi_2 \simeq \sin^2 \kappa T_1 $ and in  the interval $\kappa \geq y >2/T$
we can substitute $\cos^2 \varphi/2$ by it mean value $1/2$.
As a result we have within logarithmic accuracy
\begin{eqnarray}
&& F_2= \int_{0}^{\infty} \frac{y dy}{(1+y)^2}
\left(\frac{\kappa_0^2}{(1+\kappa_0^2+y)} \right)^2
\Phi_2(y) \simeq \int_{1}^{\kappa}\frac{dy}{y}\Phi_2(y)
\nonumber \\
&& \simeq \sin^2 \kappa T_1 \ln \frac{2}{T}+2\sin^2 \frac{\kappa T_1}{2}
\ln \frac{\kappa T}{2} \vartheta(\kappa T-1).
\label{53}
\end{eqnarray}
The function $F_2$ vanishes in the points
$\kappa T_1 \simeq \kappa_0^2 T_1=2\pi n$, the corresponding photon
energies are
\begin{equation}
\omega_{(2n)}=\frac{\omega_0^2 l_1}{4\pi n} \equiv \frac{\omega_1}{n},
\quad \omega_1=\frac{T_c}{2\pi}\frac{\omega_p^2}{\omega_c},\quad
n=1,2\ldots
\label{54}
\end{equation}
In the points $\kappa T_1 \simeq \pi(2n+1)$ the function $F_2$ has
minimums which depend on the distance between plates $l_2$
\begin{equation}
\omega_{(2n+1)} \simeq \frac{2\omega_1}{2n+1},
\quad F_2 \simeq 2 \ln \frac{\kappa T}{2}=2 \ln\left(\frac{\kappa T_1}{2}
\frac{l}{l_1} \right) \simeq 2\ln \left[\left(n+\frac{1}{2} \right)
\pi \frac{l}{l_1} \right],~l=l_1+l_2
\label{55}
\end{equation}
The function $F_2$ has maximums in the points where $\sin \kappa T_1 \simeq 1
(\kappa T_1 \simeq \pi (m+1/2))$ and in these points
\begin{equation}
\omega^{(m)} \simeq \frac{4\omega_1}{2m+1},\quad F_2 \simeq \ln \kappa,
\label{56}
\end{equation}
i.e. as the values of the function $F_2$ as well as the positions of
the maximums of $F_2$ are independent of the distance between plates in the
wide interval $l_2 \geq l_1$.

Now we perform similar analysis for arbitrary $N$. In the region
$y < 2/(NT) \gg 1$ one has
\[
\Phi_N \simeq \sin^2 \frac{N\kappa T_1}{2},
\]
and in the interval $\kappa > y > 2/T$ the phase $\varphi$ varies fast and
the function
\[
\displaystyle{\frac{\sin^2(N\varphi/2)}
{\sin^2(\varphi/2)}}=
|1+\exp(-i\varphi)+\exp(-2i\varphi)+\ldots|^2
\]
can be substituted by its mean value $N$. In this interval
$\Phi_N \simeq N \sin^2(\kappa T_1/2)$. In the intermediate region
$2/T > y > 2/(NT)$ the phases $\varphi_1$ and $\varphi$ are approximately
equal and the function $\Phi_N \simeq \sin^2 (N\varphi/2)$ oscillates fast
and can be substituted by it mean value $1/2$.
Taking this results into account we find performing the integration
over $y$
\begin{eqnarray}
&& F_N= \int_{0}^{\infty} \frac{y dy}{(1+y)^2}
\left(\frac{\kappa-1}{\kappa+y} \right)^2
\Phi_N(y) \simeq \int_{1}^{\kappa}\frac{dy}{y}\Phi_N(y)
\nonumber \\
&& \simeq \sin^2 \frac{N \kappa T_1}{2} \ln \frac{2}{NT}+\frac{1}{2}\ln N+
N\sin^2 \frac{\kappa T_1}{2}
\ln \frac{\kappa T}{2}.
\label{57}
\end{eqnarray}
It follows from this formula that positions of the minimums in the points
$\omega = \omega_{(2n)}$ are independent of $N$, while the minimums
in the points $\omega = \omega_{(2n+1)}$  are disappearing when the
value $N$ increases and for enough large $N$ the function $F_N$ has
maximums in this points. The case $N \gg 1$ was considered in detail
in our recent paper \cite{14}.

We will discuss now the results of numerical calculations given in
Figs.1,2. The formulas (\ref{31}), (\ref{33}) and (\ref{49}), (\ref{50})
were used respectively. The spectral curves of energy loss were obtained for
the case of two gold plates with the thickness $l_1=11.5~\mu m$ with different
gaps $l_2$ between plates. The initial energy of electrons is $25$~GeV.
The characteristic parameters for this case are:
\begin{eqnarray}
&& \omega_c \simeq 240~{\rm MeV},\quad T_c \simeq 2.9,\quad b^{-1} \simeq 3.3,
\quad \omega_{th} \simeq 80~{\rm MeV},
\nonumber \\
&& \omega_{p} \simeq 3.9~{\rm MeV},\quad \omega_{1} \simeq 30~{\rm keV},
\quad \frac{T}{T_1} \equiv k =\frac{l_1+l_2}{l_1} = 3,5,7,9,11.
\label{58}
\end{eqnarray}
At $\omega > 80~$MeV the radiation process occurs independently
from each plate according with theory of the LPM effect \cite{8}.
The interference pattern appears at $\omega < 80~$MeV where the
formation length is longer than the thickness of one plate and the radiation
process depends on the distance between plates $T$. According to
Eqs.(\ref{42b}), (\ref{43}) the curves 1-5  have minimums
at $\omega \simeq \pi \omega_{th}/k~(T=\pi)$ which are outside of
Fig.1 and will be discussed below. In accord with the above analysis
the spectral curves in Fig.1 have the maximums at photon
energies $\omega \simeq \omega_{th}/k~(T=1)$.
These values (in MeV) are $\omega \simeq 27, 16, 11, 9, 7$
for curves $1, 2,3, 4, 5$ respectively.
At further decrease of $\omega(T)$ the spectral curves diminish according
to Eq.((\ref{42})) and attain the minimum at
$\omega_{min} = \omega_{th}/(k(1+2/b))~(T(1+2/b)=1)$.
The corresponding values (in MeV) are $\omega \simeq 3.5, 2, 1.4, 1.2$
for curves $1, 2, 3, 4$. The least value
of $\omega_{min} \simeq 1~$MeV has the curve $5$. However, one has to take
into account that at $\omega \leq 1.5~$MeV ($\kappa_0^2=p_c^2 \simeq 2/b$)
the contribution of the transition radiation becomes significant.
Starting from $\omega \leq 0.6~$MeV the contribution of the transition
radiation dominates. The spectral curves for the transition radiation
in Fig.2 increase for $\omega < 0.2~$MeV  as $(\kappa T_1)^2\ln 2/T~
(\kappa T_1 \ll 1)$ according to Eq.(\ref{53}) and attain the maximal value
at $\omega^{(0)}=4\omega_1=0.12~$MeV (see Eq.(\ref{56})). The height of the
spectral curves at this point within the logarithmic accuracy is independent
of $T$ and roughly the same for all the curves. The minimums of the
spectral curves are disposed at $\omega_{(1)}=2\omega_1 \simeq 0.06~$MeV
according to Eq.(\ref{55}) from which it follows that in this point
for larger $T$ the spectral curve is higher. The next maximum is situated
in $\omega^{(1)}=4\omega_1/3=0.04~$MeV. At $\omega_{(2)}=\omega_1=0.03~$MeV
the spectral curves have the absolute minimum according to Eq.(\ref{53}).
At further decrease of $\omega$ the higher harmonics appear:
maximum at the point $\omega^{(2)}=4\omega_1/5=0.024~$MeV,
the minimum at $\omega_{(3)}=2\omega_1/3=0.02~$MeV etc.

The approach developed in this paper is applicable in the interval of photon
energies where effects of the polarization of a medium are essential.
It includes also the soft part of the LPM effect. For the case given
in Fig.1 our results are given up to $\omega_{max} \sim 20~$MeV.
On the other hand, in \cite{12} the hard part of the LPM effect spectrum
was analyzed where one can neglect the effects of
the polarization of a medium ($\omega > 5~$MeV for the mentioned case).
Although in our paper and in \cite{12}
the different methods are used, the results obtained in overlapping
regions are in a quite reasonable agreement among themselves.

It is interesting to discuss behavior of the spectral curves obtained in
\cite{12} from the point of view of our results. We consider the low
value $l_1~(T_c$
is not very large) and a situation when corrections to the value $b$
in (\ref{48}) which neglected in \cite{12} are less than $20\%$. This leads
to the difference in results less than $10\%$.
We concentrate on the case of gold
target with the total thickness $Nl_1=0.7\%~L_{rad}$. The case $N=1$
where $T_c=5.8,~b^{-1}=7.3,~\omega_c \simeq 240~$Mev, $\omega_{th}=
4\omega_c/(T_c(1+T_c)) \simeq 24~$MeV is considered
in detail in our paper \cite{8}. The curves in figures in \cite{12}
are normalized on the Bethe-Heitler probability of radiation, i.e. they
measured in units $\alpha r_2 T_c/(3\pi)$. In the region where our
results are applicable $\omega < \omega_{th} \simeq 24~$MeV in Fig.2$~(G=0)$
of \cite{12} one can see plateau the ordinate of which is $10\%$ less
than calculated according (\ref{30}). The case of two plates
($T_c \simeq b^{-1} \simeq 3,~\omega_c \simeq 240~$MeV,~$\omega_{th}
\simeq \omega_c/T_c = 80~$MeV) is given
in Fig.3 of \cite{12}. The lengths of the gaps are the same as in our
Fig.1, except $k=9$. The positions and ordinates of the minimums and
the maximums in the characteristic points
($\omega \simeq \pi \omega_{th}/k$ for minimums and
$\omega \simeq \omega_{th}/k~(T=1)$ for maximums, see above) as well as
behavior of the spectral curves is described quite satisfactory by
our formulas (see e.g. asymptotic Eqs.(\ref{40})-(\ref{42b})).

In the case of four plates
($T_c \simeq b^{-1} \simeq 1.5,~\omega_{th}
\simeq \omega_c/T_c = 160~$MeV, Fig.4 of \cite{12}) the value $b$ is not
enough small and we can do the qualitative analysis only.
The curves in this figure correspond to the values $k=3, 5, 9, 13$ according
to Eqs.(\ref{40})-(\ref{42b}), (\ref{44}). For $k=13$ we have from
(\ref{42b}):the first maximum is at $\omega \sim \omega_{th}/k \simeq 12~$MeV,
the first minimum is at $\omega \sim \pi \omega_{th}/k \simeq 40~$MeV,
the next maximum is at $\omega \sim 2\pi \omega_{th}/k \simeq 80~$MeV,
the next minimum is at $\omega \sim 3\pi \omega_{th}/k \simeq 115~$MeV,
but it is rather obscure because of large value of $T=3\pi$. The accordance
of these estimates of the characteristic points with the curve
in the figure is quite satisfactory.

Let us note in conclusion that for observation of the interference pattern
at large values $b$ which is described by Eq.(\ref{34}) (see Fig.3) one needs
to use very thin plates ($l_1=\alpha L_{rad}/(2\pi b)$). Since the
interference manifests itself in the terms $\propto 1/b$ only, the value
$b$ should not be very large. It seems at first sight
that in this situation one can
use the light elements, e.g. lithium, for which the radiation length
$L_{rad}$ is large. However, these elements have low charge of nucleus $Z$
and low density and because of this the LPM effect begins at
$\omega < \omega_c$ which is close to $\omega_p=\omega_0 \gamma$
where effects of the polarization of a medium are essential.
So, one can expect that the optimal situation will be for elements
in the middle of the periodic table of the elements.
For example, for $Ge$ one has $L_{rad}=2.30~$cm, $\omega_0=44~$eV, and
at energy $\varepsilon=25~$GeV, $\omega_c=34~$MeV,
$\omega_{th}=b\omega_c=100~$MeV, $\omega_p=2.2~$MeV.
Taking $b=3$ we have $l_1=8.9~\mu$m, $\omega_1=6.95~$keV. In this
situation there is rather wide interval of photon energies
$\omega_p < \omega < \omega_{th}$ where the
interference can be observed.

If one wants to observe the interference pattern G(T) given by Fig.3
in the wide interval of $T:~ T_{min} \leq T \leq T_{max},~ T_{min} \ll 1,
~T_{max} \gg 1$, then one has to take into account
that $\omega=\omega_{th}T_1=\omega_{th}T/k~(k=1+l_2/l_1)$. In this
situation one has
\begin{eqnarray}
&& \omega_{max} = \frac{T_{max}\omega_{th}}{k} < \omega_{th},
\quad k > T_{max},
\nonumber \\
&& \omega_{min} = \frac{T_{min}\omega_{th}}{k} > \omega_{p},
\quad  k < \frac{\omega_{th}}{\omega_p} T_{min}.
\label{59}
\end{eqnarray}
Acceptable parameters are $T_{max} \sim 10,~T_{min} \sim 1/5$. For
$T > 10$ the amplitude of oscillation is very small, while for
$T \geq 0.2$ we have very distinct first maximum. So we have
$k \simeq l_2/l_1 \simeq 10$ for the case considered.

\vspace{0.3cm}
{\bf Acknowledgments}

\vspace{0.2cm}
This work
was supported in part by the Russian Fund of Basic Research under Grant
98-02-17866.

\newpage

\setcounter{equation}{0}
\appendix
\Alph{equation}

\section{Appendix}

We start calculation of $g_n({\bf p}',{\bf p})$ (\ref{15}) with lower terms
$n=2,3$. Direct application of formula of the type Eq.(\ref{14})
can be written in the form
\begin{equation}
g_n({\bf p}',{\bf p})
 =\frac{b}{\pi d_n} \exp \left\{-\frac{b}{d_n} \left[
\left({\bf p}'^2+{\bf p}^2 \right)\left(d_n-d_{n-1} \right)-
2{\bf p}{\bf p}'\right]\right\},
\label{a1}
\end{equation}
where $d_1, d_2, d_3$ are given by Eq.(\ref{17}). For arbitrary $n$ we will
use the mathematical induction method. Let (\ref{a1}) is valid for $n$,
than for $n+1$ we have from definition (\ref{15})
\begin{eqnarray}
&& g_{n+1}({\bf p}',{\bf p}) = \left(\frac{b}{\pi} \right)^2 \frac{1}{d_n}
\int_{}^{}\exp \left\{-\frac{b}{d_n}\left[\left({\bf p}'^2+{\bf p}_n^2 \right)
\left(d_n-d_{n-1} \right)-2{\bf p}_n{\bf p}'\right] \right\}
\nonumber \\
&& \times \exp\left[-b\left({\bf p}_n-{\bf p} \right)^2-i{\bf p}_n^2 T
\right] d^2p_n
\label{a2}\end{eqnarray}
The integral here is of the type Eq.(\ref{14}), so we have
\begin{equation}
g_{n+1}({\bf p}',{\bf p})
 =\frac{b}{\pi d_n} \frac{b}{\beta_{n+1}}
\exp\left(\frac{\alpha_n^2}{4\beta_{n+1}} \right)
\exp \left[-\frac{b}{d_n}
\left(d_n-d_{n-1} \right){\bf p}'^2-
b{\bf p}^2\right],
\label{a3}
\end{equation}
where
\[
\beta_{n+1}=b\left(\frac{d_n-d_{n-1}}{d_n}+2\alpha-1 \right),\quad
\alpha_n^2=4b^2\left(\frac{{\bf p}'^2}{d_n^2}+2\frac{{\bf p}{\bf p}'}{d_n}
+{\bf p}^2 \right).
\]
In order that formula (\ref{a1}) will be valid for $n+1$ the following
equalities have to be fulfilled (they obtained from comparison of
expressions Eqs.(\ref{a1}) and (\ref{a3}) at corresponding combinations
of momenta)
\begin{equation}
\beta_{n+1}=\frac{b}{d_n} d_{n+1},\quad
d_n^2=1+d_{n-1}d_{n+1},\quad d_{n+1}=2\alpha d_{n}-d_{n-1}.
\label{a4}
\end{equation}
We consider the recursion relation somewhat more general than the
last relation (\ref{a4}):
\begin{equation}
D_{n+1}=a D_{n}-b^2D_{n-1},
\label{a5}
\end{equation}
with the initial conditions $D_2=a,~D_3=a^2-b^2$. We will find
the explicit form of $D_n$ using Okunev's method. The quadratic
equation $z^2-az+b^2=0$ has the roots
\begin{equation}
z_1=\frac{a}{2}+A,\quad z_2=\frac{a}{2}-A,\quad A=\sqrt{\frac{a^2}{4}-b^2}.
\label{a6}
\end{equation}
Substituting these roots into Eq.(\ref{a5}) we can write
\begin{equation}
D_{n+1}=\left(z_1+z_2 \right) D_{n}-z_1z_2D_{n-1}.
\label{a7}
\end{equation}
From this relation we have
\begin{eqnarray}
&& D_{n+1}-z_1D_n=z_2\left(D_{n}-z_1D_{n-1} \right)
=\ldots=z_2^{n-2}\left(D_{3}-z_1D_{2} \right),
\nonumber \\
&& D_{n+1}-z_2D_n=z_1\left(D_{n}-z_2D_{n-1} \right)
=\ldots=z_1^{n-2}\left(D_{3}-z_2D_{2} \right).
\label{a8}\end{eqnarray}
Multiplying the first relation by $z_2$ and the second relation by $z_1$
and subtracting one from another we obtain
\begin{eqnarray}
\hspace{-14mm}&& D_{n+1}=\frac{1}{2A} \Bigg[\left(\frac{a}{2}+A \right)^{n-1}
\left(a^2-b^2-a\left(\frac{a}{2}-A \right) \right)
\nonumber \\
\hspace{-14mm}&&-\left(\frac{a}{2}-A \right)^{n-1}
\left(a^2-b^2-a\left(\frac{a}{2}+A \right) \right) \Bigg]
 = \frac{1}{2A} \left[\left(\frac{a}{2}+A \right)^{n+1}
-\left(\frac{a}{2}-A \right)^{n+1}\right],
\label{a9}\end{eqnarray}
where $A$ is defined in Eq.(\ref{a6}).
If $a=2\alpha,~b=1$ we have from the last formula Eq.(\ref{16}).
The second recursion relation of Eq.(\ref{a4}) is also satisfied by
Eq.(\ref{16}).

Note that $n$-lines determinant
\[
D_{n+1}=
\left| \begin{array}{lllcll}
a & b & 0 & \cdots & 0 & 0 \\
b & a & b & \cdots & 0 & 0 \\
0 & b & a & \cdots & 0 & 0 \\
\multicolumn{6}{c}\dotfill \\
0 & 0 & 0 & \cdots & a & b \\
0 & 0 & 0 & \cdots & b & a \\
\end{array}
\right|
\]
gives the recursion relation (\ref{a5}) and so Eq.(\ref{a9})
gives value of this determinant.

\newpage

{\bf Figure captions}

\vspace{15mm}
\begin{itemize}

\item {\bf Fig.1} The energy losses spectrum
$\displaystyle{\frac{d\varepsilon}{d\omega}}$
in units $\displaystyle{\frac{2\alpha}{\pi}}$,
in the target consisting
of two gold plates with thickness $l_1=11.5~\mu m$
for the initial electrons energy $\varepsilon$=25~GeV .
\begin{itemize}
\item Curve 1 is for distance between plates $l_2=2l_1$;
\item Curve 2 is for distance between plates $l_2=4l_1$;
\item Curve 3 is for distance between plates $l_2=6l_1$;
\item Curve 4 is for distance between plates $l_2=8l_1$;
\item Curve 5 is for distance between plates $l_2=10l_1$.
\end{itemize}

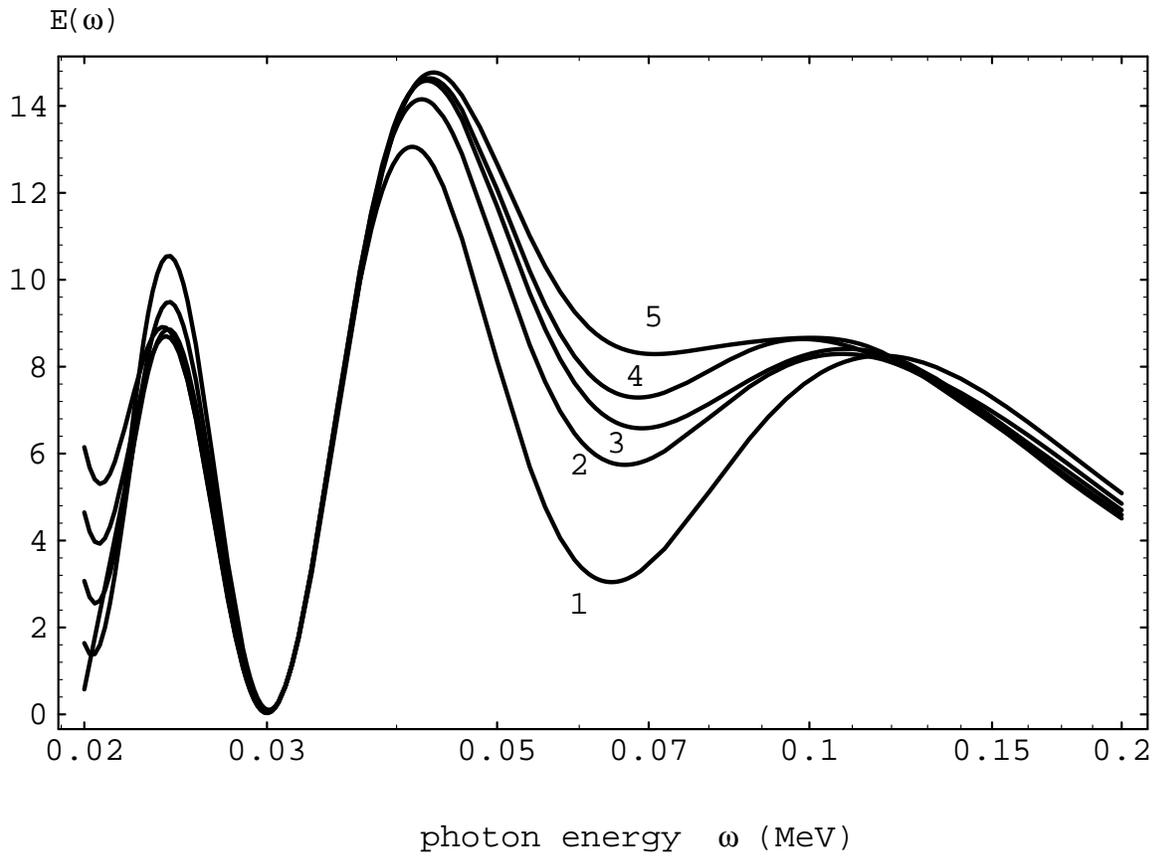
\item {\bf Fig.2} The same as in Fig.1
$\displaystyle{E(\omega)=\frac{d\varepsilon}{d\omega}}$
in soft part of the spectrum where transition radiation contributes only.
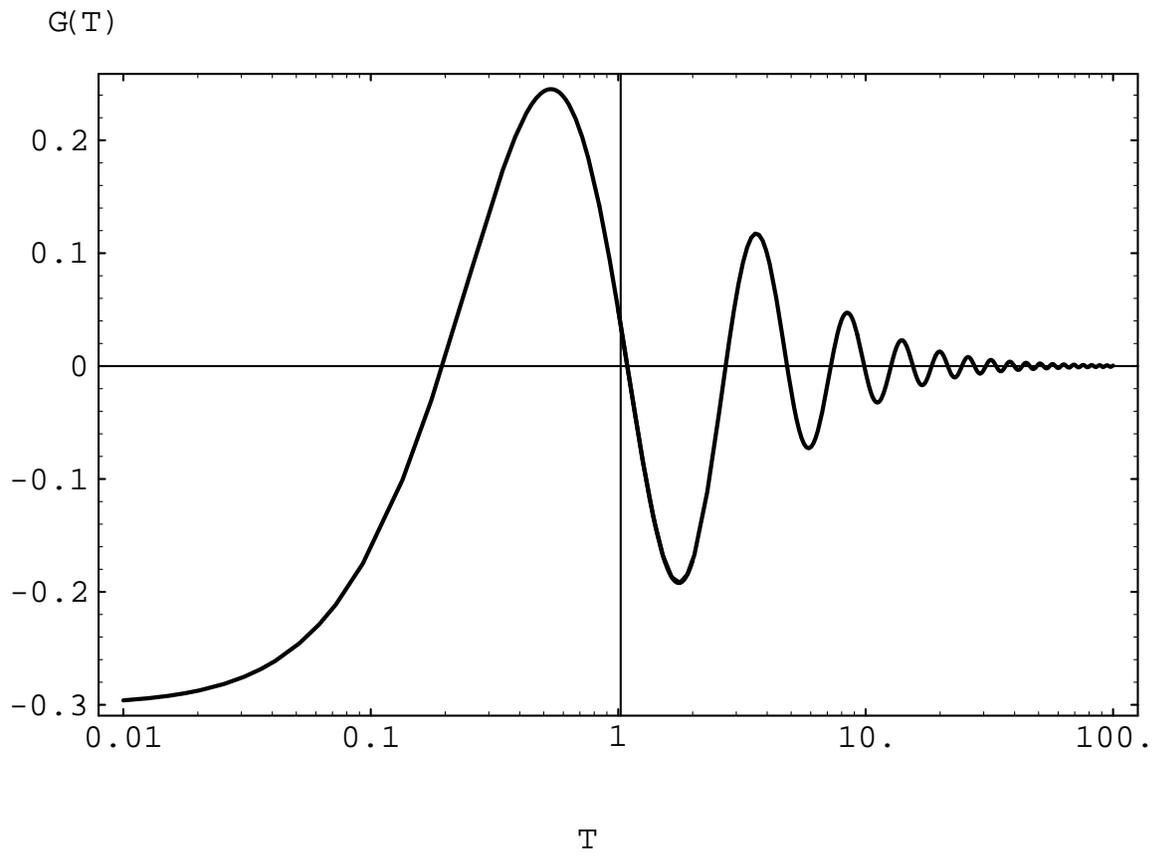
\item {\bf Fig.3} The function $G(T)$ (\ref{34}).
\end{itemize}

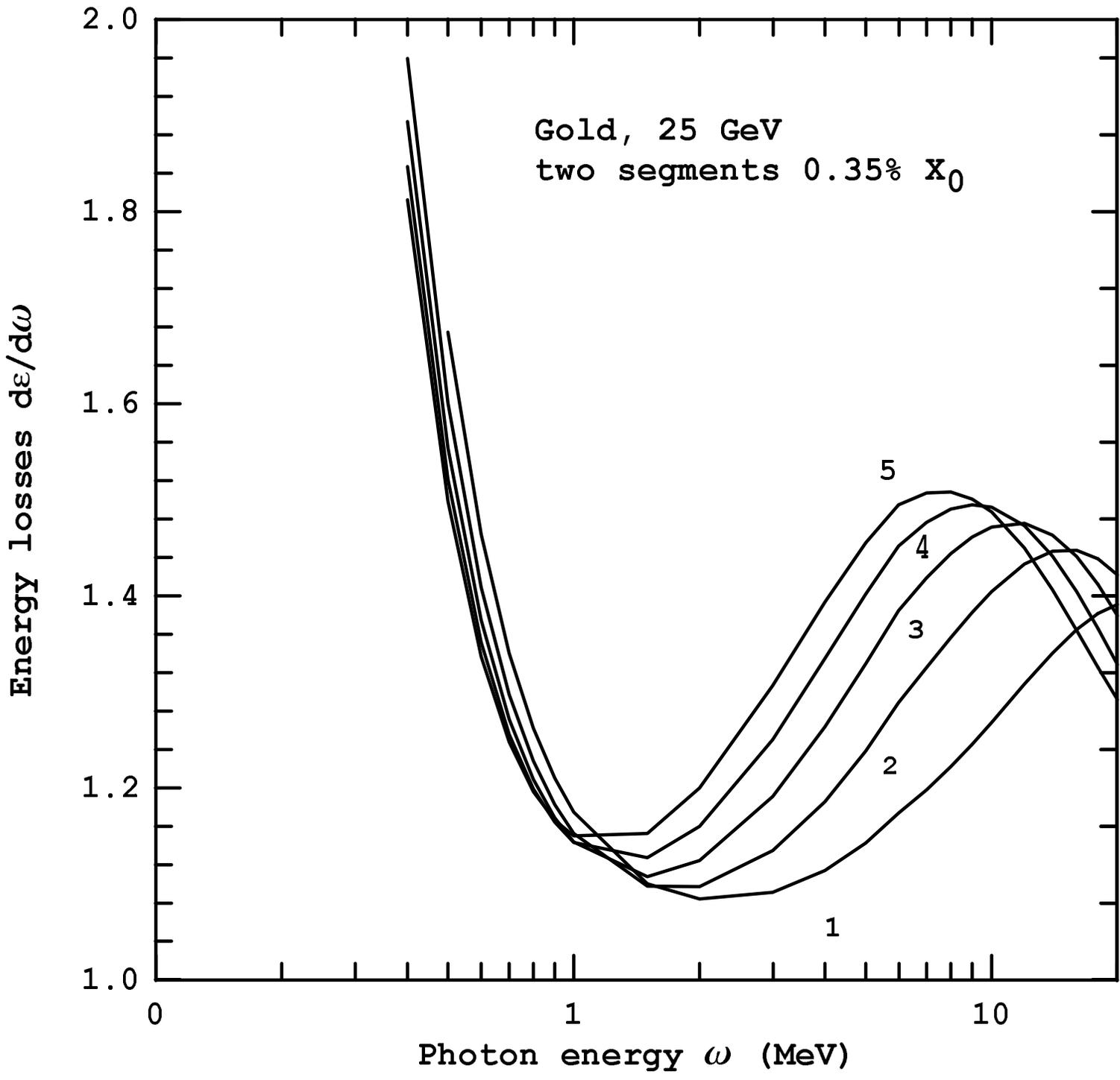
\end{document}